\documentclass[preprint]{iucr}              
     \journalcode{B}              

\usepackage{amsmath}
\usepackage{graphicx}

\usepackage{multirow}
\usepackage{tabularx}
\usepackage{array,afterpage,longtable}
\usepackage{lscape} 
\newcolumntype{Y}{>{\centering\arraybackslash}X}

\begin{document}                  



\title{Effect of pressure on the order -- disorder phase transitions of $B$-cations in $AB'_{1/2}B''_{1/2}$O$_3$ perovskites}
\shorttitle{Effect of pressure on order -- disorder transitions in $AB'_{1/2}B''_{1/2}$O$_3$}


\cauthor{N. V.}{Ter-Oganessian}{teroganesyan@sfedu.ru}

\author{V. P.}{Sakhnenko}

\aff{Institute of Physics, Southern Federal University, 344090 Rostov-on-Don, \country{Russia}}











\maketitle                        

\begin{synopsis}
The effect of pressure on the order -- disorder phase transitions of $B$-cations in $AB'_{1/2}B''_{1/2}$O$_3$ perovskites is theoretically studied and estimates are made for certain compositions. In many cases pressure can significantly increase or decrease the order-disorder transition temperature, which provides another way to manipulate cation ordering.
\end{synopsis}

\begin{abstract}
Perovskite-like oxides $AB'_{1/2}B''_{1/2}$O$_3$ may experience different ordering degrees of $B$-cations, that can be varied by suitable synthesis conditions or post-synthesis treatment. In this work the earlier proposed statistical model of order-disorder phase transitions of $B$-cations is extended to account for the effect of pressure. Depending on composition, pressure is found to either increase or decrease the order-disorder phase transition temperature. The change of transition temperature due to pressure in many cases reaches several hundreds of kelvin at pressures accessible in laboratory, which may significantly change the atomic ordering degree. The work is intended to help determining how pressure influences the degree of atomic ordering and stimulate research of the effect of pressure on atomic order-disorder phase transitions in perovskites.
\end{abstract}


\section{Introduction}

Many natural and synthetic crystal systems contain several types of atoms distributed over equivalent positions in the crystal lattice. The changes of thermodynamic state parameters can result in variation of the atomic distribution up to the appearance of atomically ordered structures, which has pronounced effect on the physical properties of crystals. Atomic order-disorder phase transitions are particularly pronounced in alloys, on which the main theoretical studies are focused~\cite{Miracle_2017}. Similar phenomena are also characteristic of a number of classes of oxides, e.g., perovskites, spinels, etc.~\cite{King_2010,Talanov_2014}. Various crystal structures with different atomic ordering degrees appearing in them often possess qualitatively different static and dynamic macroscopic properties. Crystals with the same chemical formula, but with different ordering degrees, can be ferroelectric, antiferroelectric, or can show various magnetic ordering patterns, demonstrating qualitatively different diffusive character of the respective phase transitions, whereas in dynamics they can experience relaxor or spin glass properties~\cite{Isupov_2003}.

The synthesis of crystals with various degrees of atomic ordering is mainly achieved by different temperature and time regimes of annealing. In some cases such synthesis and annealing is performed under application of hydrostatic pressure. The prominent examples are CuAu alloy~\cite{Asaumi_CuAu_1975}, PbSc$_{1/2}$Nb$_{1/2}$O$_3$~\cite{Zhu_PSN_2008}, and PbIn$_{1/2}$Nb$_{1/2}$O$_3$~\cite{Ohwada_PIN_2008}. It is usually assumed that atomic ordering generally results in decrease in volume, which forms the basis for thermodynamic account of the influence of pressure on atomic order-disorder transitions~\cite{Hazen_1996,Angel_Pressure_effect_1999}. Thus, it is concluded that application of pressure initially results in an increase of the order-disorder phase transition temperature and, respectively, of the  cation order. The effect of further increase of pressure may be not so straightforward since, for example, in statics the ordered and disordered phases may have different bulk moduli resulting in reduction of the volume difference between phases with pressure, whereas in dynamics higher pressure may significantly hamper the atomic diffusion.

One of the most representative classes of polyatomic oxides is the perovskite family $AB'_{1/2}B''_{1/2}$O$_3$. Large number of works is devoted to this family, which are summarized and analyzed in the recent review by~\citeasnoun{Vasala_2015}. In most of the available literature high pressure is used for synthesis of perovskites only in case when the perovskite structure can not be obtained under normal pressure. This is the case, for example, for PbFe$_{1/2}$Sb$_{1/2}$O$_3$~\cite{Raevski_PFS_2013}, BiFe$_{1/2}$Sc$_{1/2}$O$_3$ \cite{Khalyavin_Bi2FeScO6_2014}, Na$M_{1/2}$Te$_{1/2}$O$_3$ ($M$ = Ti, Sn)~\cite{Park_Na2TiTeO6_1999}, and CaCu$_3$Ga$_2M_2$O$_{12}$ ($M$ = Sb, Ta)~\cite{Byeon_QuadrPerov_2003}. Interestingly, PbFe$_{1/2}$Sb$_{1/2}$O$_3$ and Na$M_{1/2}$Te$_{1/2}$O$_3$ ($M$ = Ti, Sn) possess an ordered arrangement of atoms at the $B$-site, in CaCu$_3$Ga$_2M_2$O$_{12}$ ($M$ = Sb, Ta) both the $A$- and the $B$-sublattices are ordered, whereas BiFe$_{1/2}$Sc$_{1/2}$O$_3$ remains disordered.

In turn, SrFe$_{1/2}$Sb$_{1/2}$O$_3$ and SrFe$_{1/2}$Re$_{1/2}$O$_3$ were obtained under normal pressure with the atomic ordering degrees $s=0.5$ and 0.75, respectively, whereas the high-pressure synthesized samples were almost completely ordered with $s\approx1$~\cite{Marysko_SFS_2017,Retuerto_Sr2FeReO6_2009}. Furthermore, it was possible to obtain PbFe$_{1/2}$Sb$_{1/2}$O$_3$ with different degrees of atomic ordering $s$=0.17 to 0.96 by tuning the synthesis temperature and time under high-pressure conditions~\cite{Raevski_PFS_2016}.

In~\citeasnoun{Sakhnenko_2018} we introduced a theoretical model for calculation of the order-disorder phase transition temperature of $B$-cations in $AB'_{1/2}B''_{1/2}$O$_3$ and calculated the dependence of some crystallographic properties of crystals on the degree of atomic ordering $s$. Among external influences, the hydrostatic pressure is distinguished by the fact that it can substantially change interatomic distances and, therefore, the cation-anion interactions, which play the principal role in cationic ordering process. In this work we extend our model to account for the effect of hydrostatic pressure on the atomic order-disorder phenomena in $AB'_{1/2}B''_{1/2}$O$_3$.

\section{Thermodynamic potential and account for pressure}

In this work we closely follow the statistical model of atomic ordering in the $B$-sublattice introduced by~\citeasnoun{Sakhnenko_2018} and the details can be found therein. Here we briefly review the model. The degree of 1:1 rock-salt type atomic ordering of the $B'$ and $B''$ cations in $AB'_{1/2}B''_{1/2}$O$_3$ splitting the $B$-sublattice into two interpenetrating sublattices denoted by single and double primes can be written as
\begin{equation}
s=\frac{N'_{B'}-N'_{B''}}{N_{\rm tot}/2},
\end{equation}
where $N_{\rm tot}$ is the total number of simple $AB$O$_3$ perovskite unit cells, $N'_{B'}$ and $N'_{B''}$ are the numbers of atoms $B'$ and $B''$ in one of the introduced sublattices, respectively.

The elastic energies of the $B'$-O and $B''$-O bonds denoted, respectively, $U_{B'{\rm O}(B'')}$ and $U_{B''{\rm O}(B')}$ can be written as
\begin{align}
U_{B'{\rm O}(B'')}&=\frac{k_{B'}}{2}\left(\frac{a}{2}+u-l_{B'}\right)^2,\label{eq:Energy_UB1B2}\\
U_{B''{\rm O}(B')}&=\frac{k_{B''}}{2}\left(\frac{a}{2}-u-l_{B''}\right)^2,
\end{align}
where $a$ is the reduced cubic cell parameter corresponding to one $AB'_{1/2}B''_{1/2}$O$_3$ formula unit of a generally distorted perovskite cell and $u$ is the oxygen displacement along the cubic $B$-O-$B$ direction, i.e., displacement of the oxygen out of the (100) plane towards one of the $B$-cations. The elastic energies $U_{B'{\rm O}(B')}$ and $U_{B''{\rm O}(B'')}$ corresponding to the case of the same $B$ cations in neighboring cells as well as the elastic energy $U_{A{\rm O}}$ of the $A$-O bond take the forms
\begin{align}
U_{B'{\rm O}(B')}&=\frac{k_{B'}}{2}\left(\frac{a}{2}-l_{B'}\right)^2,\\
U_{B''{\rm O}(B'')}&=\frac{k_{B''}}{2}\left(\frac{a}{2}-l_{B''}\right)^2,\\
U_{A{\rm O}}&=\frac{k_{A}}{2}\left(\frac{a\sqrt{2}}{2}-l_{A}\right)^2.\label{eq:Energy_UA}
\end{align}
Here $l_{A}$, $l_{B'}$, and $l_{B''}$ are unstrained equilibrium bond lengths $A$ -- O, $B'$ -- O, and $B''$ -- O, whereas $k_{A}$, $k_{B'}$, and $k_{B''}$  are their stiffnesses, respectively. The stiffnesses are related to the atomic valences $n_{A}$ and $n_{B}$ of the ions $A$ and $B$, respectively, through
\begin{equation}
k_{A}=\frac{n_{A}}{4}\gamma,\qquad k_{B}=n_{B}\gamma,\label{eq:k_n_relation}
\end{equation}
where $\gamma\approx70$~N/m is a constant~\cite{Sakhnenko_2018}. The set of unstrained equilibrium bond lengths $l_{A}$, $l_{B'}$, and $l_{B''}$ for majority of cations was determined by comparison of the model for predictions of reduced lattice constants of perovskites with experimental literature data of a broad range of compounds~\cite{Sakhnenko_1972}. The full set with updated values for some elements is available in~\citeasnoun{Sakhnenko_2018}.

It should be noted that the modeling of cation-anion interactions by quasielastic forces also includes the Coulombic interaction, taking part in formation of the respective potential energy. In the linear approximation used in our work this energy can be considered as the first (and the strongest) term in expansion of the commonly used model interaction potentials (e.g. Morse, Lennard-Jones, etc.) with respect to deviation from the equilibrium distance assumed in the work. This is the basic assumption of the model. Electrostatic interaction of ions beyond nearest neighbors is not taken into account in our work.

In order to account for the effect of pressure $P$ we extend the model introduced in~\citeasnoun{Sakhnenko_2018} by constructing the thermodynamic potential
\begin{equation}
G=E-TS+PV\label{eq:thermodynamic_potential},
\end{equation}
where the internal energy is written as
\begin{align}
E=&\frac{3}{2}N_{\rm tot}\left(U_{B'{\rm O}(B')}+U_{B'{\rm O}(B'')}+U_{B''{\rm O}(B')}+U_{B''{\rm O}(B'')}\right)\nonumber\\
 &+\frac{3}{2}N_{\rm tot}s^2\left(-U_{B'{\rm O}(B')}+U_{B'{\rm O}(B'')}+U_{B''{\rm O}(B')}-U_{B''{\rm O}(B'')}\right)+12N_{\rm tot}U_{A{\rm O}},\label{eq:InternalEnergy}
\end{align}
the entropy in the Gorsky-Bragg-Williams approximation is given by
\begin{equation}
S=-kN_{\rm tot}\left(\frac{1+s}{2}\ln\frac{1+s}{2}+\frac{1-s}{2}\ln\frac{1-s}{2}\right),
\end{equation}
where $k$ is the Boltzmann constant and
\begin{equation}
V=N_{\rm tot}a^3,
\end{equation}
is volume.


All contributions to the internal energy~(\ref{eq:InternalEnergy}), as follows from equations~(\ref{eq:Energy_UB1B2}) -- (\ref{eq:Energy_UA}), change in different ways with decrease of the lattice parameter $a$ under pressure. For stretched $B$-O and $A$-O bonds, the decrease of $a$ results in decrease of the quasielastic bond energy, whereas the contribution of compressed bonds increases. It is the complex balance of these interactions that should determine the change in the ordering temperature when the pressure is applied.

The system of equations
\begin{equation}
\frac{\partial G}{\partial a}=0,\qquad \frac{\partial G}{\partial u}=0,\label{eq:dGdA_dGdu}
\end{equation}
allows determining the equilibrium values of $a$ and $u$ for the given values of the order parameter $s$ and pressure $P$. To the first order in $P$ they can be represented as
\begin{align}
a=a_0+a_1P,\\
u=u_0+u_1P,
\end{align}
where
\begin{eqnarray}
a_0(s)&=&2\left[ 8 \sqrt{2} k_{A} l_{A} (k_{B'}+k_{B''})+ k_{B'} k_{B''}(l_{B'}+l_{B''})(3+s^2)\right.\nonumber\\
 & &\left. + (k_{B'}^2 l_{B'}+k_{B''}^2 l_{B''})(1-s^2) \right]
\Delta^{-1},\\
u_0(s)&=&4\left\{ k_{B'}k_{B''}\left(l_{B'}-l_{B''}\right) +2k_{A}\left[\sqrt{2}l_{A}\left(k_{B''}-k_{B'}\right) \right.\right.\nonumber\\
 & &\left.\left. + 2\left(k_{B'} l_{B'}-k_{B''} l_{B''}\right)\right]\right\}
\Delta^{-1},\\
a_1(s)&=&-\frac{8 a_0^2 \left(k_{B'}+k_{B''}\right)}{\Delta},\\
u_1(s)&=&\frac{4 a_0^2 \left(k_{B'}-k_{B''}\right)}{\Delta},\label{eq:u1}
\end{eqnarray}
and
\begin{equation}
\Delta=k_{B'}^2+6 k_{B'} k_{B''}+k_{B''}^2+16 k_{A} (k_{B'}+k_{B''})-(k_{B'}-k_{B''})^2s^2.
\end{equation}

From the equation $\partial G/\partial s=0$ one can obtain the temperature dependence of the order parameter in the vicinity of the phase transition
\begin{equation}
s^2(T)=\frac{3(T_{\rm od}-T)}{T},
\end{equation}
where, taking into account equations~(\ref{eq:k_n_relation}) and the relation
\begin{equation}
u=\frac{a(k_{B''}-k_{B'})+2k_{B'}l_{B'}-2k_{B''}l_{B''}}{2(k_{B'}+k_{B''})}
\end{equation}
following from equations~(\ref{eq:dGdA_dGdu}),
\begin{eqnarray}
T_{\rm od}&=&\frac{3(n_{B'}+n_{B''})}{2k}\gamma u^2(s=0) \label{eq:T_od_of_u}\\
&=&\frac{3(2l_{B'}n_{B'}-2l_{B''}n_{B''}+a(s=0)(n_{B''}-n_{B'}))^2}{8k(n_{B'}+n_{B''})}\gamma.
\end{eqnarray}

According to equation~(\ref{eq:T_od_of_u}) the behavior of $T_{\rm od}$ with pressure is determined by the pressure dependence of $u(s=0)$. In our calculations we assume that $n_{B'}\leq n_{B''}$, which means that $u_1(s)\leq0$. Thus, $T_{\rm od}$ increases [decreases] with pressure if $u_0(s=0)<0$ [$u_0(s=0)>0$]. Positive $u$ means that the oxygen atom is displaced towards $B''$, whereas negative value means that it is displaced towards $B'$. The sign of $u$ is largely determined by whether $l_{B'}$ is bigger or smaller than $l_{B''}$, i.e., by the relation between the sizes of cations $B'$ and $B''$. As follows from table 1 in~\citeasnoun{Sakhnenko_2018} the sign of $u_0(s=0)$ correlates with the lattice parameter difference between the disordered and ordered states $a(s=0)-a(s=1)$, i.e., the conditions of $u_0(s=0)<0$ [$u_0(s=0)>0$] and $a(s=0)-a(s=1)>0$ [$a(s=0)-a(s=1)<0$] are equivalent, which agrees with the general conclusion that pressure promotes atomic ordering if $a(s=0)>a(s=1)$ and impedes if $a(s=0)<a(s=1)$. Furthermore, it follows from equation~(\ref{eq:u1}) that when the ions $B'$ and $B''$ have the same valence, pressure does not influence the ordering temperature $T_{\rm od}$, which agrees with the fact that in this case $a(s=0)=a(s=1)$~\cite{Sakhnenko_2018}.

\begin{figure}
\caption{Schematic diagram showing the behavior of $T_{\rm od}$ of (a) $AB'^{3+}_{1/2}B''^{5+}_{1/2}$O$_3$ ($A$=Ca, Sr, Ba, Pb; $B'$=Al, Ga, Cr, Fe, Sc, In, Yb; $B''$=Sb, Nb, Ta) and (b) $AB'^{2+}_{1/2}B''^{4+}_{1/2}$O$_3$ ($A$=Ho, Nd, La, Bi; $B'$=Ni, Mg, Zn, Mn; $B''$=Mn, Ru, Ti, Sn, Hf, Zr) with pressure. Open and closed circles and stars give positions of the perovskites in the $l_{B'}$--$l_{B''}$ plane depending on $B'$ and $B''$ cations. Four solid lines corresponding to $u_0(s=0)$ delimit the regions of $dT_{\rm od}/dP>0$ and $dT_{\rm od}/dP<0$ for the particular $A$-cation.}
\label{fig:PhaseDiag}
\includegraphics[height=8.85cm]{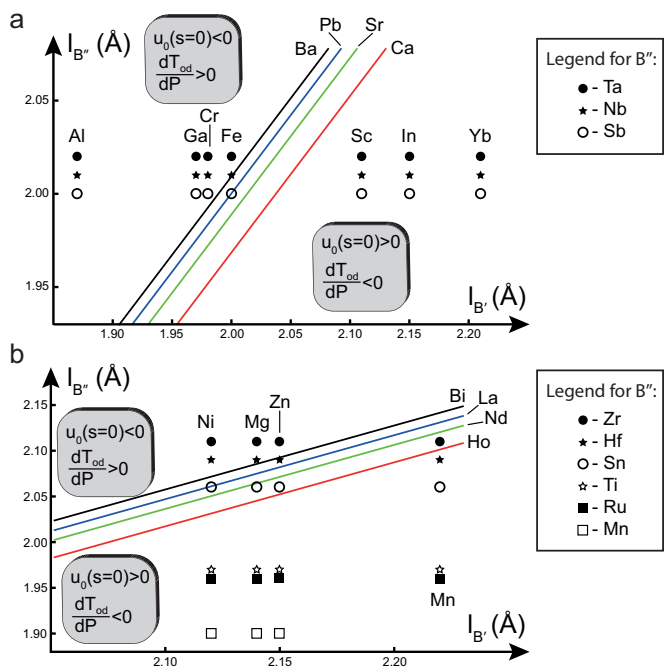}
\end{figure}

The behavior of $T_{\rm od}$ with pressure for $AB'^{3+}_{1/2}B''^{5+}_{1/2}$O$_3$ perovskites, which most easily undergo the atomic order-disorder phenomena, is determined by the sign of [see table 1 in~\citeasnoun{Sakhnenko_2018}]
\begin{equation}
u_0(s=0)=\frac{2\sqrt{2}l_A+21l_{B'}-25l_{B''}}{47}.
\end{equation}
Thus, for the given $A$-cation one can plot the regions of $u_0(s=0)>0$ and $u_0(s=0)<0$ in the $l_{B'}$ -- $l_{B''}$ plane as shown in Fig.~\ref{fig:PhaseDiag}. Roughly, it appears that $u_0(s=0)>0$ for $B'^{3+}$ cations larger than $B''^{5+}$ ones, e.g., $B'$=Sc, In, Yb and $B''$=Nb, Ta, Sb, which means that pressure suppresses atomic ordering. In turn, for smaller $B'^{3+}$ cations like Al$^{3+}$ or Ga$^{3+}$ pressure stimulates atomic ordering. The farther the point is from the delimiting line in Fig.~\ref{fig:PhaseDiag}, the higher $T_{\rm od}$. Table~\ref{tab:3-5_perovskites} lists the order-disorder temperatures $T_{\rm od}$ of $AB'^{3+}_{1/2}B''^{5+}_{1/2}$O$_3$ perovskites and their pressure dependence.

It was reported that PbFe$_{1/2}$Sb$_{1/2}$O$_3$ and SrFe$_{1/2}$Sb$_{1/2}$O$_3$ synthesized under pressure of 6~GPa show almost complete order in the $B$-sublattice~\cite{Marysko_Sr2FeSbO6_2017,Misjul_2013}. The former compound can only be obtained under pressure, whereas the latter compound can also be obtained without pressure with the ordering degree of $s=0.5$. According to our results $T_{\rm od}$ of $A$Fe$_{1/2}$Sb$_{1/2}$O$_3$ is rather low similar to the perovskites $A$Fe$_{1/2}$Nb$_{1/2}$O$_3$ and $A$Fe$_{1/2}$Ta$_{1/2}$O$_3$, which indeed do not show $B$-cation ordering. However, the special behavior of $A$Fe$_{1/2}$Sb$_{1/2}$O$_3$ perovskites is related to strong covalency of Sb--O bonds~\cite{Goodenough_ABO3_1973,Mizoguchi_NaSbO3_2004}. The Sb$^{5+}$ ion has a filled $d^{10}$ shell, which does not participate in $\pi$-bonding. Thus, the Fe$^{3+}$ and Sb$^{5+}$ ions tend to order, because the 180$^\circ$ linear Sb--O--Sb configuration of $\sigma$-bonds is unfavorable, whereas the empty $d^0$ shells of Nb$^{5+}$ and Ta$^{5+}$ are capable of forming $\pi$-bonds, which lifts this restriction for Nb- and Ta-containing perovskites. Therefore, in order to calculate $T_{\rm od}$ for Sb-containing perovskites using our model one has to introduce additional energy required for the formation of Sb--O--Sb configuration.

According to the present model in most of the $AB'^{2+}_{1/2}B''^{6+}_{1/2}$O$_3$ ($A$ = Ca, Sr, Ba, Pb; $B'$ = Mg, Ca, Cr, Mn, Fe, Co, Zn; $B''$ = W, Re) perovskites, which are actively studied in literature, the ordering temperature is very high exceeding the melting temperature and confirming the fact that these perovskites are ordered already at synthesis. However, our results show that pressure lowers their $T_{\rm od}$, which possibly can be used as a way for obtaining their disordered analogues. Thus, in table~\ref{tab:2-6_perovskites} we list $AB'^{2+}_{1/2}B''^{6+}_{1/2}$O$_3$ ($A$ = Ca, Sr, Ba, Pb; $B'$ = Mg, Ca, Cr, Mn, Fe, Co, Zn; $B''$ = W, Re) perovskites with $T_{\rm od}$ not exceeding 5000~K under zero pressure. Under pressure of 10~GPa their $T_{\rm od}$ decreases by ~1000 - 2000~K possibly becoming lower than the synthesis temperature, given the estimative character of the data.

The perovskites $AB'^{2+}_{1/2}B''^{4+}_{1/2}$O$_3$ with a trivalent cation at the $A$-site can experience order-disorder phenomena according to the present model similar to the case of $AB'^{3+}_{1/2}B''^{5+}_{1/2}$O$_3$ perovskites discussed above, since for many of them $T_{\rm od}$ lies in the range 1000 -- 3000~K. In table~\ref{tab:2-4_perovskites} we list representative compounds of this family with $A$=Bi, La, Nd, Ho; $B'$=Zn, Mg, Mn, Ni; and $B''$=Ti, Sn, Zr, Ru, Mn. It can be noted that many Sn- and Zr-containing perovskites possess low $T_{\rm od}$. La$_2$MgSnO$_6$ and Nd$_2$MgSnO$_6$ were found to have an ordered $B$-cation arrangement despite the fact that their $T_{\rm od}$ values are 34 and 3~K, respectively~\cite{Babu_LaMgSn_2007,Babu_LaNdMgSn_2008}. However, similar to the Sb$^{5+}$ case above, Sn$^{4+}$ has considerable covalency of Sn-O bonding and filled $4d$ shell, which does not enable $\pi$-bonds that can stabilize the Sn$^{4+}$--O--Sn$^{4+}$ linkage. Thus, such Sn$^{4+}$--O--Sn$^{4+}$ linkage has additional energy that needs to be accounted for in order to describe cation ordering in $AB_{1/2}$Sn$_{1/2}$O$_3$ perovskites.

There is a limited number of experimental works in literature on $A^{3+}B^{2+}_{1/2}$Zr$_{1/2}$O$_3$ perovskites. According to our calculations the compounds with $B$=Zn, Mg, Mn, and Ni have $T_{\rm od}$ below 1000~K, which lowers the probability of obtaining them with ordered $B$-cation arrangement. However, in several cases pressure can significantly increase $T_{\rm od}$ as follows from table~\ref{tab:2-4_perovskites}. La$_2$CaZrO$_6$ has been obtained with the ordered Ca--Zr arrangement~\cite{Levin_2002} in agreement with $T_{\rm od}$=2922~K according to the present model. Interestingly, $T_{\rm od}$=1989~K at a pressure of 10~GPa, which makes it possible to increase atomic disorder in this compound.

Finally it has to be noted that (i) many transition metals with partially filled $d$-shells can experience high, low, or intermediate spin states, and (ii) $B'$/$B''$ pairs may possess variable oxidation states, e.g., Fe$^{2+}$/Mo$^{6+}$ and Fe$^{3+}$/Mo$^{5+}$. Both phenomena have strong influence on the sizes of cations and, thus, on the order-disorder transitions, which can be accounted for in the calculations. In the present work we considered all the elements in the high-spin states and with fixed valences.

In the introduction part we noted that pressure substantially influences the atomic ordering in SrFe$_{1/2}$Re$_{1/2}$O$_3$, which is a rare example of a perovskite that was experimentally obtained both with and without application of pressure. SrFe$_{1/2}$Re$_{1/2}$O$_3$ is among the compounds with frequently occurring variable oxidation states of Fe. Indeed, it follows from the literature data that Fe ions with mixed oxidation states of +2.5 and +3 are observed in this compound~\cite{Blanco_Sr2FeReO6_2009,Retuerto_Sr2FeReO6_2010}. From our model it follows that SrFe$^{3+}_{1/2}$Re$^{5+}_{1/2}$O$_3$ and SrFe$^{2+}_{1/2}$Re$^{6+}_{1/2}$O$_3$ have very different ordering temperatures $T_{\rm od}$ of 0.2 and 7253~K, respectively, at zero pressure. Thus, the atomic ordering degree is determined by the oxidation states of Fe and Re at the synthesis or annealing temperatures, which are difficult to determine. Moreover, the pressure influence will also depend on pressure dependence of the oxidation states. From our point of view, the theoretical description of atomic ordering in such compounds deserves separate consideration.

\section{Discussion of the model and its limitations}

The present approach to the analysis of atomic ordering in perovskites is based on the modeling of thermodynamic potential~(\ref{eq:thermodynamic_potential}) in the Gorsky-Bragg-Williams approximation. Here we review the simplifications of our model and its applicability to the solution of the atomic ordering problem. The entropy $S$ includes only the configurational part expressed through the long-range order parameter. Therefore, $S$ has a universal form for all $AB'_{1/2}B''_{1/2}$O$_3$ compounds regardless of their composition. In order to determine the internal energy of the crystal lattice at different atomic ordering degrees one needs to calculate the energy of interaction of nearest cation-anion neighbors, which gives the main contribution to $E$. For this purpose we use the model of unstrained cation-anion bond lengths~\cite{Sakhnenko_1972}, deviation from which in a complex crystal lattice is characterized by quasi-elastic energy taken into account in harmonic approximation. The respective stiffness coefficient $k$ between the cation and the anion $X$~(\ref{eq:k_n_relation}) is found using the empirical formula obtained by~\citeasnoun{Gordy_1946} from the analysis of infrared spectra of molecules
\begin{equation}
k=\eta\frac{n}{Z}\left(\frac{\chi\chi_X}{l^2}\right)^{3/4},
\end{equation}
where $n$ is the cation valence, $Z$ is its coordination number, $\chi$ and $\chi_X$ are electronegativities of the cation and the anion, respectively, $l$ is the cation-anion equilibrium distance, and $\eta$ is a constant.

In the previous work the acceptability of internal energy calculation in the quasi-elastic approximation was demonstrated by relation between the reduced lattice parameters of the perovskite-like compounds with the chemical formulas $ABX_3$, $A'BX_3$, $AB'X_3$, $A'B'X_3$, where $A$, $A'$ and $B$, $B'$ are pairs of cations with equal valences, and $X$ is oxygen or halogen~[equation~(4) in~\citeasnoun{Sakhnenko_2018}]. According to our model the reduced lattice parameters are less sensitive to the stiffness coefficients of $A$-O and $B$-O bonds than the bond energy. Here for the assessment of the approximation~(\ref{eq:k_n_relation}) for elastic constants we make use of comparison of the properties of oxides and fluorides. In our model the bulk modulus $K$ and the reduced lattice parameter $a$ of perovskites are related by equation~(16) in~\citeasnoun{Sakhnenko_2018}
\begin{equation}
Ka=\frac{(n_A+n_B)\gamma}{6},
\end{equation}
from which it follows that the ratio of $(Ka)_{ox}$ for oxides to that for fluorides $(Ka)_{fl}$ is given by
\begin{equation}
\frac{(Ka)_{ox}}{(Ka)_{fl}}=\frac{(n_A+n_B)_{ox}}{(n_A+n_B)_{fl}}\cdot\frac{\gamma_{ox}}{\gamma_{fl}}\approx2,\label{eq:oxy_fluo_ratio}
\end{equation}
where $\gamma_{ox}/\gamma_{fl}\approx0.9$ is calculated taking into account the electronegativities of oxygen and fluoride. The bulk modulus and the reduced lattice constants for several fluorides are given in supporting information to this work, whereas for perovskites one can use the data collected in Table~2 of~\citeasnoun{Sakhnenko_2018}. Averaging the data for oxides and fluorides we obtain $(Ka)_{ox}\approx71.8$~N m$^{-1}$ and $(Ka)_{fl}\approx29.6$~N m$^{-1}$, whereas their ratio~(\ref{eq:oxy_fluo_ratio}) is equal to 2.4 in fairly good agreement with the theoretical estimate.

In order to be able to compare the results for perovskites with different compositions, the model is developed for a cubic lattice, which corresponds to the high-temperature phases of the majority of perovskites. It is in this phase that the ordering process begins in most of them. On the other hand, structural phase transitions occurring at decreasing temperature can influence both the entropy and internal energy. However, in most of the perovskites the cationic displacements are significantly smaller than the displacements of oxygens.  Furthermore, most structural phase transitions in perovskites are due to rotations of oxygen octahedra, which corresponds to displacements of oxygens normal to the line connecting two neighboring $B$-cations. Therefore, the changes of cation-anion distances due to such phase transitions are small, which allows assuming that our calculations performed for cubic crystals will be also acceptable for crystals experiencing structural phase transitions.

In the present work the temperatures of atomic ordering phase transitions and the degrees of atomic ordering are determined for states in thermodynamic equilibrium, whose achievement is to a great extent governed by the features of cation diffusion. We assume that the degree of ordering is not strongly influenced by structural transformations of crystals, since in many cases structural phase transitions are of second order or close to them transitions of first order, which results in small changes of cation-anion distances and the respective energies in comparison to the cubic state. However, the kinetic processes may be more sensitive to such changes and can substantially differ for compounds with different compositions. At the same time, the temporal characteristics of diffusion processes, especially the difference of high and low temperature diffusion, allow obtaining stable at low temperatures compounds with varying degrees of cationic order and, consequently, substantially different macroscopic properties. In turn, external pressure applied during synthesis or annealing processes, that may also have large nonhydrostatic contribution, can influence both the lattice symmetry and diffusion processes, which may have certain impact on both the ordering temperature and kinetic processes.

\section{Conclusions}

In the present work we extended the statistical model of $B$-site cation ordering in $AB'_{1/2}B''_{1/2}$O$_3$ perovskites to account for the effect of pressure~\cite{Sakhnenko_2018}. Depending on specific composition, pressure can either increase or decrease the order-disorder phase transition temperature $T_{\rm od}$ and the temperature change $\Delta T_{\rm od}$ reaches several hundreds of kelvin at 10~GPa. In many cases this can result in significant shift of $T_{\rm od}$ leading to respective changes in atomic ordering degree depending on the relation of $T_{\rm od}$ to the synthesis or post-synthesis treatment temperatures. To the best of our knowledge pressure is rarely used in the synthesis of perovskites if the desired compound can be obtained at ambient pressure. Therefore, we hope that (i) the present model can stimulate research of the effect of pressure on cation ordering in perovskites, and (ii) in many cases may help synthesizing perovskites with the degree of atomic ordering higher or lower than that obtained at ambient pressure.

In conclusion we would like to point out the substantial difference of the problem of 1:1 $A$-site cation ordering from the ordering in the $B$-sublattice studied in the present work. As was already pointed out by~\citeasnoun{Knapp_2006} the absence of $A'_{1/2}A''_{1/2}B$O$_3$ compounds with the rock-salt type ordering of the $A$-sublattice is due to the geometric constraint related to the structure of the cubic phase. In the statistical model of cation-anion bonds, which we study, this constraint is due to increase of the lattice energy at 1:1 $A$-site ordering because of impossibility of elastic stress relaxation by the anion displacement, which preserves a centrosymmetric position in the 1:1 $A$-site ordered structure. However, in complex perovskites with composition $A'_{1/2}A''_{1/2}B'_{1/2}B''_{1/2}$O$_3$ upon the $B$-sublattice ordering the centrosymmetry of the oxygen position is violated, which may result in lower elastic energies of $A'$-O-$A'$ and $A''$-O-$A''$ bonds, than of the $A'$-O-$A''$ bonds corresponding to the disordered state of the $A$-sublattice. Cation ordering in such perovskites can be thought of as of a two-stage process: the $B$-sublattice should order first, followed by the possible ordering of the $A$-sublattice at lower temperatures. The estimation of the values of thermodynamic state parameters, such as temperature and pressure, at which the ordering of the $B$-sublattice begins, can be made using the results of the present work assuming averaged characteristics of the $A$-cations ($l_{A}$ and $k_{A}$). Consistent theoretical treatment of this problem with the description of mutual influence of the $A$- and $B$-sublattice orderings can be performed using the scheme proposed by~\citeasnoun{Sakhnenko_2018} and in the present work.


\pagebreak

\begin{longtable}{l|c|c|c|c|c}      
\caption{Reduced lattice constant $a(0)$ at $s=0$ and $P=0$ and the difference $\Delta a=a(0)-a(1)$ between the disordered and ordered ($s=1$) cases, order-disorder phase transition temperatures $T_{\rm od}^{P=0~{\rm GPa}}$ and $T_{\rm od}^{P=10~{\rm GPa}}$ at $P=0$ and 10~GPa, respectively, and their difference $\Delta T_{\rm od}=T_{\rm od}^{P=10~{\rm GPa}}-T_{\rm od}^{P=0~{\rm GPa}}$ for $A_2B'^{3+}B''^{5+}$O$_6$ ($A$=Ca, Sr, Ba, Pb; $B'$=Al, Ga, Cr, Fe, Sc, In, Yb; $B''$=Ta, Nb, Sb) perovskites.\label{tab:3-5_perovskites}}
\\
\hline
Formula & $a(0)$, \AA & $\Delta a$, $10^{-2}$ \AA & $T_{\rm od}^{P=0~{\rm GPa}}$, K  & $T_{\rm od}^{P=10~{\rm GPa}}$, K  & $\Delta T_{\rm od}$, K \\
\hline
\multicolumn{6}{c}{ $A$ = Ca}\\
Ca$_2$AlTaO$_6$ & 3.81 & 1.5 & 4448 & 5411 & 963 \\
Ca$_2$AlNbO$_6$ & 3.80 & 1.4 & 3911 & 4814 & 903 \\
Ca$_2$YbSbO$_6$ & 3.99 & -1.3 & 3616 & 2766 & -850 \\
Ca$_2$AlSbO$_6$ & 3.79 & 1.3 & 3409 & 4252 & 843 \\
Ca$_2$YbNbO$_6$ & 3.99 & -1.3 & 3134 & 2344 & -790 \\
Ca$_2$YbTaO$_6$ & 4.00 & -1.2 & 2687 & 1957 & -730 \\
Ca$_2$InSbO$_6$ & 3.95 & -0.9 & 1538 & 1012 & -526 \\
Ca$_2$InNbO$_6$ & 3.96 & -0.8 & 1230 & 763 & -467 \\
Ca$_2$GaTaO$_6$ & 3.86 & 0.7 & 1013 & 1514 & 501 \\
Ca$_2$InTaO$_6$ & 3.97 & -0.7 & 956 & 550 & -406 \\
Ca$_2$CrTaO$_6$ & 3.87 & 0.6 & 803 & 1256 & 453 \\
Ca$_2$GaNbO$_6$ & 3.86 & 0.6 & 766 & 1206 & 440 \\
Ca$_2$ScSbO$_6$ & 3.93 & -0.6 & 639 & 322 & -317 \\
Ca$_2$CrNbO$_6$ & 3.86 & 0.5 & 585 & 978 & 393 \\
Ca$_2$GaSbO$_6$ & 3.85 & 0.5 & 554 & 934 & 380 \\
Ca$_2$FeTaO$_6$ & 3.88 & 0.5 & 457 & 813 & 356 \\
Ca$_2$ScNbO$_6$ & 3.94 & -0.5 & 446 & 190 & -256 \\
Ca$_2$CrSbO$_6$ & 3.85 & 0.5 & 402 & 734 & 332 \\
Ca$_2$FeNbO$_6$ & 3.87 & 0.4 & 297 & 592 & 295 \\
Ca$_2$ScTaO$_6$ & 3.94 & -0.4 & 288 & 92 & -196 \\
Ca$_2$FeSbO$_6$ & 3.87 & 0.3 & 171 & 406 & 235 \\
\multicolumn{6}{c}{ $A$ = Sr}\\
Sr$_2$YbSbO$_6$ & 4.07 & -1.5 & 4704 & 3686 & -1018 \\
Sr$_2$YbNbO$_6$ & 4.08 & -1.4 & 4152 & 3196 & -956 \\
Sr$_2$YbTaO$_6$ & 4.09 & -1.3 & 3634 & 2741 & -893 \\
Sr$_2$AlTaO$_6$ & 3.89 & 1.3 & 3392 & 4281 & 889 \\
Sr$_2$AlNbO$_6$ & 3.89 & 1.2 & 2926 & 3751 & 825 \\
Sr$_2$AlSbO$_6$ & 3.88 & 1.1 & 2494 & 3257 & 763 \\
Sr$_2$InSbO$_6$ & 4.04 & -1.1 & 2272 & 1595 & -677 \\
Sr$_2$InNbO$_6$ & 4.05 & -1.0 & 1894 & 1278 & -616 \\
Sr$_2$InTaO$_6$ & 4.05 & -0.9 & 1550 & 997 & -553 \\
Sr$_2$ScSbO$_6$ & 4.02 & -0.8 & 1137 & 680 & -457 \\
Sr$_2$ScNbO$_6$ & 4.02 & -0.7 & 875 & 479 & -396 \\
Sr$_2$ScTaO$_6$ & 4.03 & -0.6 & 646 & 314 & -332 \\
Sr$_2$GaTaO$_6$ & 3.95 & 0.5 & 547 & 947 & 400 \\
Sr$_2$CrTaO$_6$ & 3.96 & 0.4 & 396 & 746 & 350 \\
Sr$_2$GaNbO$_6$ & 3.94 & 0.4 & 370 & 708 & 338 \\
Sr$_2$CrNbO$_6$ & 3.95 & 0.4 & 248 & 535 & 287 \\
Sr$_2$GaSbO$_6$ & 3.94 & 0.3 & 227 & 503 & 276 \\
Sr$_2$FeTaO$_6$ & 3.97 & 0.3 & 167 & 416 & 249 \\
Sr$_2$CrSbO$_6$ & 3.94 & 0.3 & 134 & 359 & 225 \\
Sr$_2$FeNbO$_6$ & 3.96 & 0.2 & 77 & 263 & 186 \\
Sr$_2$FeSbO$_6$ & 3.95 & 0.1 & 21 & 145 & 124 \\
\multicolumn{6}{c}{ $A$ = Ba}\\
Ba$_2$YbSbO$_6$ & 4.16 & -1.7 & 5934 & 4737 & -1197 \\
Ba$_2$YbNbO$_6$ & 4.17 & -1.6 & 5312 & 4179 & -1133 \\
Ba$_2$YbTaO$_6$ & 4.17 & -1.5 & 4725 & 3656 & -1069 \\
Ba$_2$InSbO$_6$ & 4.13 & -1.3 & 3149 & 2309 & -840 \\
Ba$_2$InNbO$_6$ & 4.13 & -1.2 & 2701 & 1924 & -777 \\
Ba$_2$AlTaO$_6$ & 3.98 & 1.1 & 2479 & 3283 & 804 \\
Ba$_2$InTaO$_6$ & 4.14 & -1.1 & 2287 & 1575 & -712 \\
Ba$_2$AlNbO$_6$ & 3.97 & 1.0 & 2083 & 2822 & 739 \\
Ba$_2$ScSbO$_6$ & 4.10 & -0.9 & 1779 & 1169 & -610 \\
Ba$_2$AlSbO$_6$ & 3.96 & 0.9 & 1721 & 2395 & 674 \\
Ba$_2$ScNbO$_6$ & 4.11 & -0.9 & 1446 & 901 & -545 \\
Ba$_2$ScTaO$_6$ & 4.12 & -0.8 & 1148 & 667 & -481 \\
Ba$_2$GaTaO$_6$ & 4.04 & 0.3 & 223 & 513 & 290 \\
Ba$_2$CrTaO$_6$ & 4.04 & 0.3 & 131 & 368 & 237 \\
Ba$_2$GaNbO$_6$ & 4.03 & 0.2 & 116 & 341 & 225 \\
Ba$_2$CrNbO$_6$ & 4.04 & 0.2 & 53 & 225 & 172 \\
Ba$_2$GaSbO$_6$ & 4.02 & 0.2 & 44 & 204 & 160 \\
Ba$_2$FeTaO$_6$ & 4.05 & 0.1 & 20 & 150 & 130 \\
Ba$_2$FeSbO$_6$ & 4.04 & -0.1 & 15 & 15 & 0 \\
Ba$_2$CrSbO$_6$ & 4.03 & 0.1 & 10 & 117 & 107 \\
Ba$_2$FeNbO$_6$ & 4.05 & 0.0 & 0 & 65 & 65 \\
\multicolumn{6}{c}{ $A$ = Pb}\\
Pb$_2$YbSbO$_6$ & 4.12 & -1.6 & 5370 & 4254 & -1116 \\
Pb$_2$YbNbO$_6$ & 4.13 & -1.5 & 4779 & 3726 & -1053 \\
Pb$_2$YbTaO$_6$ & 4.14 & -1.5 & 4222 & 3234 & -988 \\
Pb$_2$AlTaO$_6$ & 3.94 & 1.2 & 2867 & 3710 & 843 \\
Pb$_2$InSbO$_6$ & 4.09 & -1.2 & 2742 & 1975 & -767 \\
Pb$_2$AlNbO$_6$ & 3.93 & 1.1 & 2440 & 3218 & 778 \\
Pb$_2$InNbO$_6$ & 4.09 & -1.1 & 2325 & 1621 & -704 \\
Pb$_2$AlSbO$_6$ & 3.93 & 1.0 & 2047 & 2762 & 715 \\
Pb$_2$InTaO$_6$ & 4.10 & -1.0 & 1942 & 1302 & -640 \\
Pb$_2$ScSbO$_6$ & 4.06 & -0.9 & 1476 & 936 & -540 \\
Pb$_2$ScNbO$_6$ & 4.07 & -0.8 & 1174 & 697 & -477 \\
Pb$_2$ScTaO$_6$ & 4.08 & -0.7 & 907 & 494 & -413 \\
Pb$_2$GaTaO$_6$ & 4.00 & 0.4 & 349 & 690 & 341 \\
Pb$_2$CrTaO$_6$ & 4.00 & 0.3 & 231 & 520 & 289 \\
Pb$_2$GaNbO$_6$ & 3.99 & 0.3 & 211 & 488 & 277 \\
Pb$_2$CrNbO$_6$ & 4.00 & 0.3 & 122 & 347 & 225 \\
Pb$_2$GaSbO$_6$ & 3.98 & 0.2 & 108 & 321 & 213 \\
Pb$_2$FeTaO$_6$ & 4.02 & 0.2 & 68 & 252 & 184 \\
Pb$_2$CrSbO$_6$ & 3.99 & 0.2 & 48 & 209 & 161 \\
Pb$_2$FeNbO$_6$ & 4.01 & 0.1 & 17 & 137 & 120 \\
Pb$_2$FeSbO$_6$ & 4.00 & 0.0 & 0 & 57 & 57 \\
\hline
\end{longtable}


\pagebreak

\begin{longtable}{l|c|c|c|c|c}      
\caption{Reduced lattice constant $a(0)$ at $s=0$ and $P=0$ and the difference $\Delta a=a(0)-a(1)$ between the disordered and ordered ($s=1$) cases, order-disorder phase transition temperatures $T_{\rm od}^{P=0~{\rm GPa}}$ and $T_{\rm od}^{P=10~{\rm GPa}}$ at $P=0$ and 10~GPa, respectively, and their difference $\Delta T_{\rm od}=T_{\rm od}^{P=10~{\rm GPa}}-T_{\rm od}^{P=0~{\rm GPa}}$ for some $A_2B'^{2+}B''^{6+}$O$_6$ perovskites.\label{tab:2-6_perovskites}}
\\
\hline
Formula & $a(0)$, \AA & $\Delta a$, $10^{-2}$ \AA & $T_{\rm od}^{P=0~{\rm GPa}}$, K  & $T_{\rm od}^{P=10~{\rm GPa}}$, K  & $\Delta T_{\rm od}$, K \\
\hline
Ca$_2$ZnReO$_6$ & 3.82 & -3.6 & 4893 & 3044 & -1849 \\
Ca$_2$CoReO$_6$ & 3.82 & -3.6 & 4893 & 3044 & -1849 \\
Pb$_2$CrReO$_6$ & 3.92 & -3.6 & 4853 & 2929 & -1924 \\
Ba$_2$CrWO$_6$ & 3.97 & -3.6 & 4786 & 2827 & -1959 \\
Ca$_2$FeReO$_6$ & 3.81 & -3.4 & 4504 & 2743 & -1761 \\
Ca$_2$MgReO$_6$ & 3.81 & -3.4 & 4504 & 2743 & -1761 \\
Ca$_2$ZnWO$_6$ & 3.83 & -3.2 & 3775 & 2166 & -1609 \\
Ca$_2$CoWO$_6$ & 3.83 & -3.2 & 3775 & 2166 & -1609 \\
Pb$_2$CrWO$_6$ & 3.93 & -3.1 & 3740 & 2068 & -1672 \\
Sr$_2$CrReO$_6$ & 3.87 & -3.1 & 3556 & 1980 & -1576 \\
Ca$_2$FeWO$_6$ & 3.83 & -3.0 & 3434 & 1913 & -1521 \\
Ca$_2$MgWO$_6$ & 3.83 & -3.0 & 3434 & 1913 & -1521 \\
Sr$_2$CrWO$_6$ & 3.88 & -2.6 & 2614 & 1285 & -1329 \\
Ca$_2$CrReO$_6$ & 3.77 & -2.1 & 1729 & 737 & -992 \\
Ca$_2$CrWO$_6$ & 3.79 & -1.7 & 1094 & 343 & -751 \\
\hline
\end{longtable}

\pagebreak

\begin{longtable}{l|c|c|c|c|c}      
\caption{
Reduced lattice constant $a(0)$ at $s=0$ and $P=0$ and the difference $\Delta a=a(0)-a(1)$ between the disordered and ordered ($s=1$) cases, order-disorder phase transition temperatures $T_{\rm od}^{P=0~{\rm GPa}}$ and $T_{\rm od}^{P=10~{\rm GPa}}$ at $P=0$ and 10~GPa, respectively, and their difference $\Delta T_{\rm od}=T_{\rm od}^{P=10~{\rm GPa}}-T_{\rm od}^{P=0~{\rm GPa}}$ for $A_2B'^{2+}B''^{4+}$O$_6$ ($A$=Bi, La, Nd, Ho; $B'$=Ni, Mg, Zn, Mn; $B''$=Mn, Ru, Ti, Sn, Hf, Zr) perovskites.\label{tab:2-4_perovskites}}
\\
\hline
Formula & $a(0)$, \AA & $\Delta a$, $10^{-2}$ \AA & $T_{\rm od}^{P=0~{\rm GPa}}$, K  & $T_{\rm od}^{P=10~{\rm GPa}}$, K  & $\Delta T_{\rm od}$, K\\
\hline
Bi$_2$ZnMnO$_6$ & 3.96 & -1.9 & 5530 & 4318 & -1212 \\
Bi$_2$MgMnO$_6$ & 3.96 & -1.9 & 5135 & 3972 & -1163 \\
La$_2$ZnMnO$_6$ & 3.92 & -1.8 & 4938 & 3816 & -1122 \\
La$_2$MgMnO$_6$ & 3.92 & -1.8 & 4565 & 3492 & -1073 \\
Bi$_2$MnTiO$_6$ & 4.03 & -1.7 & 4390 & 3284 & -1106 \\
Bi$_2$NiMnO$_6$ & 3.95 & -1.7 & 4390 & 3324 & -1066 \\
Nd$_2$ZnMnO$_6$ & 3.89 & -1.7 & 4379 & 3345 & -1034 \\
Nd$_2$MgMnO$_6$ & 3.88 & -1.7 & 4029 & 3042 & -987 \\
La$_2$MnTiO$_6$ & 3.99 & -1.6 & 3864 & 2848 & -1016 \\
La$_2$NiMnO$_6$ & 3.91 & -1.6 & 3864 & 2886 & -978 \\
Ho$_2$ZnMnO$_6$ & 3.82 & -1.5 & 3458 & 2575 & -883 \\
Nd$_2$MnTiO$_6$ & 3.96 & -1.5 & 3372 & 2444 & -928 \\
Nd$_2$NiMnO$_6$ & 3.88 & -1.5 & 3372 & 2478 & -894 \\
Ho$_2$MgMnO$_6$ & 3.82 & -1.5 & 3147 & 2310 & -837 \\
Bi$_2$ZnRuO$_6$ & 3.99 & -1.3 & 2621 & 1797 & -824 \\
Ho$_2$MnTiO$_6$ & 3.89 & -1.3 & 2570 & 1793 & -777 \\
Ho$_2$NiMnO$_6$ & 3.81 & -1.3 & 2570 & 1822 & -748 \\
Bi$_2$MgRuO$_6$ & 3.99 & -1.3 & 2352 & 1577 & -775 \\
Bi$_2$ZnTiO$_6$ & 4.00 & -1.2 & 2241 & 1483 & -758 \\
La$_2$ZnRuO$_6$ & 3.96 & -1.2 & 2219 & 1479 & -740 \\
Bi$_2$MgTiO$_6$ & 4.00 & -1.2 & 1992 & 1283 & -709 \\
La$_2$MgRuO$_6$ & 3.95 & -1.2 & 1971 & 1280 & -691 \\
La$_2$ZnTiO$_6$ & 3.96 & -1.1 & 1870 & 1195 & -675 \\
Bi$_2$NiRuO$_6$ & 3.98 & -1.1 & 1857 & 1179 & -678 \\
Nd$_2$ZnRuO$_6$ & 3.92 & -1.1 & 1850 & 1192 & -658 \\
La$_2$MgTiO$_6$ & 3.96 & -1.1 & 1643 & 1017 & -626 \\
Nd$_2$MgRuO$_6$ & 3.92 & -1.1 & 1625 & 1014 & -611 \\
Bi$_2$NiTiO$_6$ & 3.99 & -1.0 & 1539 & 928 & -611 \\
Nd$_2$ZnTiO$_6$ & 3.93 & -1.0 & 1532 & 939 & -593 \\
La$_2$NiRuO$_6$ & 3.95 & -1.0 & 1521 & 924 & -597 \\
Nd$_2$MgTiO$_6$ & 3.92 & -1.0 & 1328 & 781 & -547 \\
Ho$_2$ZnRuO$_6$ & 3.86 & -0.9 & 1270 & 753 & -517 \\
La$_2$NiTiO$_6$ & 3.95 & -0.9 & 1234 & 703 & -531 \\
Nd$_2$NiRuO$_6$ & 3.91 & -0.9 & 1218 & 701 & -517 \\
Ho$_2$MgRuO$_6$ & 3.85 & -0.9 & 1085 & 612 & -473 \\
Ho$_2$ZnTiO$_6$ & 3.86 & -0.8 & 1010 & 555 & -455 \\
Bi$_2$MnSnO$_6$ & 4.08 & -0.8 & 993 & 498 & -495 \\
Nd$_2$NiTiO$_6$ & 3.92 & -0.8 & 963 & 510 & -453 \\
Ho$_2$NiZrO$_6$ & 3.93 & 0.8 & 923 & 1514 & 591 \\
Ho$_2$MgTiO$_6$ & 3.86 & -0.8 & 845 & 435 & -410 \\
Ho$_2$NiRuO$_6$ & 3.84 & -0.7 & 758 & 375 & -383 \\
La$_2$MnSnO$_6$ & 4.04 & -0.7 & 751 & 338 & -413 \\
Ho$_2$MgZrO$_6$ & 3.94 & 0.7 & 624 & 1125 & 501 \\
Ho$_2$NiTiO$_6$ & 3.85 & -0.6 & 560 & 240 & -320 \\
Nd$_2$MnSnO$_6$ & 4.01 & -0.6 & 544 & 209 & -335 \\
Nd$_2$NiZrO$_6$ & 4.00 & 0.6 & 530 & 1012 & 482 \\
Ho$_2$ZnZrO$_6$ & 3.94 & 0.6 & 496 & 952 & 456 \\
La$_2$NiZrO$_6$ & 4.03 & 0.5 & 358 & 778 & 420 \\
Nd$_2$MgZrO$_6$ & 4.00 & 0.5 & 310 & 700 & 390 \\
Ho$_2$MnSnO$_6$ & 3.94 & -0.4 & 254 & 54 & -200 \\
Nd$_2$ZnZrO$_6$ & 4.01 & 0.4 & 222 & 565 & 343 \\
Bi$_2$NiZrO$_6$ & 4.07 & 0.4 & 220 & 574 & 354 \\
La$_2$MgZrO$_6$ & 4.04 & 0.4 & 183 & 507 & 324 \\
Bi$_2$ZnSnO$_6$ & 4.05 & -0.3 & 158 & 12 & -146 \\
Bi$_2$MnZrO$_6$ & 4.11 & -0.3 & 149 & 8 & -141 \\
Ho$_2$NiSnO$_6$ & 3.90 & 0.3 & 123 & 379 & 256 \\
La$_2$ZnZrO$_6$ & 4.04 & 0.3 & 117 & 393 & 276 \\
Bi$_2$MgSnO$_6$ & 4.05 & -0.3 & 98 & 1 & -97 \\
Bi$_2$MgZrO$_6$ & 4.08 & 0.3 & 89 & 345 & 256 \\
La$_2$ZnSnO$_6$ & 4.02 & -0.2 & 72 & 0 & -72 \\
La$_2$MnZrO$_6$ & 4.07 & -0.2 & 66 & 1 & -65 \\
Bi$_2$ZnZrO$_6$ & 4.08 & 0.2 & 45 & 253 & 208 \\
La$_2$MgSnO$_6$ & 4.01 & -0.2 & 34 & 10 & -24 \\
Ho$_2$MgSnO$_6$ & 3.91 & 0.2 & 32 & 199 & 167 \\
Bi$_2$NiSnO$_6$ & 4.04 & -0.1 & 20 & 20 & 0 \\
Nd$_2$ZnSnO$_6$ & 3.98 & -0.1 & 19 & 19 & 0 \\
Nd$_2$MnZrO$_6$ & 4.04 & -0.1 & 16 & 25 & 9 \\
Nd$_2$NiSnO$_6$ & 3.97 & 0.1 & 14 & 154 & 140 \\
Ho$_2$MnZrO$_6$ & 3.97 & 0.1 & 11 & 145 & 134 \\
Ho$_2$ZnSnO$_6$ & 3.91 & 0.1 & 9 & 130 & 121 \\
Nd$_2$MgSnO$_6$ & 3.98 & 0.0 & 3 & 49 & 46 \\
La$_2$NiSnO$_6$ & 4.00 & 0.0 & 0 & 71 & 71 \\
\hline
\end{longtable}




\end{document}